\documentstyle[aps,prl,preprint,graphicx,floats]{revtex}
\begin{document}
\draft
%{\wideabs{
\title{Global Optimization by  Energy Landscape Paving}
\author{Ulrich H. E. Hansmann}
\address{Department of Physics, Michigan Technological University,
Houghton, MI 49931}
\author{Luc T. Wille}
\address{Department of Physics, Florida Atlantic University, Boca Raton, 
FL 33431}
\date{\today}
\maketitle
 \ \\ \\ \\
\begin{abstract}
We introduce a novel heuristic global optimization method, 
energy landscape paving (ELP), which combines
core ideas from energy surface deformation and tabu search.
In appropriate limits, ELP reduces to existing techniques.
The approach is very general and flexible and is illustrated here
on two protein folding problems.  For these examples, the technique 
gives faster convergence
to the global minimum than previous approaches.
\end{abstract}
%}

Global optimization is one of the key issues in modern science,
technology and economy. Typical examples are 
the problem of optimal transportation routes \cite{pursula01},  
finding molecular conformations \cite{wales99,hansmann99,wille00}
or fitting experimental spectra.\cite{karle}
Consequently, much effort has been spent on designing methods 
to finding global optima.
For this purpose, the system has to be described by an objective function,
and optimality is achieved when this function reaches 
its global minimum. If  the objective function  is viewed 
as an `energy'  the optimal solution corresponds
 to the deepest minimum in the energy landscape. 
For most applications of practical interest,
competing interactions and frustration in  the system  
lead to an energy landscape with many local minima separated by
high barriers. 
Since conventional minimization techniques tend to get trapped in 
whichever local minimum they encounter first it turns out to be extremely
difficult to find the global minimum in such cases.

A general characteristic of successful  
optimization techniques is that they avoid entrapment in local minima 
and continue to explore the energy landscape for further solutions.
For instance, in  tabu search \cite{glover,cvijovic}  
the search is guided away from areas that have already been
explored in an effort to cover all important regions of the solution space.
The danger with such an approach is that it may result in slow convergence
since it does not distinguish between important and less
important regions  of the landscape.

Entrapment in local minima can also be avoided if the search is performed
in a deformed or smoothed energy landscape, for example by lowering 
diffusion barriers,\cite{besold} in stochastic tunneling \cite{HW} 
or the various generalized ensemble approaches. \cite{hansmann99} 
In the optimal case the original energy landscape is transformed in a
funnel-landscape and convergence toward the global minimum is fast.
Although they have been very successful, most of these methods require 
a considerable amount of fine-tuning or {\it a priori} information. 
Moreover, problems may exist when connecting back to the original landscape
since minima on the deformed surface may have been displaced or merged.

Here we introduce a novel approach to the global optimization problem that 
combines ideas from tabu search and energy landscape deformation. The new 
method, {\it energy landscape paving} (ELP), avoids some of the pitfalls 
of the other two and has very general applicability.
The central idea is to perform low-temperature Monte Carlo (MC) simulations, 
but with a modified energy expression designed to
steer the search away from regions that have already been explored.  
To be specific, we choose as the statistical weight for a state 
\begin{equation}
 w(\tilde{E}) = e^{-\tilde{E}/k_BT},
\end{equation}
where $T$ is a (low) temperature and $\tilde{E}$ the following 
replacement of the energy $E$:
\begin{equation}
 E \longrightarrow  \tilde{E} = E + f(H(q,t))~.
\end{equation}
In this expression, $f(H(q,t))$ is a  function of 
the histogram $H(q,t)$ in a pre-chosen  ``order parameter'' $q$. The
histogram is  updated at each MC step, hence the ``time'' 
dependence of $H(q,t)$. As a result, the search process keeps 
track of the number of prior explorations of a particular 
region in order parameter space and biases against revisiting
the same types of states.  Rather than using the  system  states
themselves in the histograms an appropriate order parameter is employed. 
This may be a ``natural'' quantity for the system under study
(such as a spheroidal deformation for a cluster) or
 the energy itself may be taken as the order parameter.  

In a regular low-temperature  simulation the probability to escape a
local minimum depends only on the height of the surrounding energy barriers.
Within ELP the weight 
of a local minimum state decreases with the time the system stays
in that minimum, and consequently the probability to escape the minimum
increases. Hence, ELP utilizes the interplay of two factors.
Given equal frequencies $H(q,t)$, the simulation will favor low energies,
thus insuring that no unphysical  high-energy
conformations are sampled. However, soon the system will
run into a local minimum. With time, ELP deforms the energy landscape 
locally  in such way that the local minimum is no longer 
favored and the system will explore higher energies. It will then either fall
in a new local minimum or walk through this high energy
region till the corresponding histogram entries all have similar frequencies.
At that point the original energy landscape is restored (that is, only shifted
by a constant (and irrelevant) factor), and the system again has a bias 
toward low energies.

ELP bears some similarities to tabu search \cite{glover,cvijovic}  
in that recently visited 
regions are not likely to be revisited immediately. 
Revisitation moves are not completely forbidden, 
but are given an exponentially lower weight compared to moves that go 
to regions with a comparable energy that have been explored less.  
With a short-term memory in the histogram and infinite cost for `forbidden' 
moves ELP becomes completely equivalent to tabu search.  On 
the other hand, ELP is also akin to an 
energy deformation approach \cite{hansmann99,besold}
in that the additional histogram parameter may be viewed as a (continuously 
changing) deformation of the energy landscape depending on the
frequency with which a particular area (as characterized by
its order parameter) has already been explored. Obviously for 
$f(H(q,t)) = f(H(q))$ the method reduces to the various generalized-ensemble 
methods.\cite{hansmann99} 
 
We have tested ELP in the context of protein folding, which 
involves the prediction of the biologically
active conformation of a protein solely from the sequence of amino acids. 
Assuming that this structure is thermodynamically stable, it is reasonable 
to identify the global-minimum conformation in the {\it free} energy at 
$T \approx 300$ K with the lowest {\it potential} energy conformation
and to choose the potential energy of the protein as an objective function.
The complexity and importance of the problem make it an ideal target for a 
test of our new optimization technique. 

As with any optimization method, ELP requires the choice of
an energy function by which the multitude of protein configurations
can be discriminated. 
%Various force fields were proposed for this purpose and 
Here, we choose the ECEPP/2 force field,\cite{EC} a commonly 
used energy function in protein simulations, as implemented in the  
computer code SMMP.\cite{SMMP}

In order to test and illustrate ELP, we concentrated on the 
structure prediction of two molecules. The first system is 
the pentapeptide Met-enkephalin,  which 
has become a frequently used benchmark model to examine new 
algorithms. We know
from previous work that the ground state of this peptide with the 
ECEPP/2 force field is given by $E_0 = -10.7~kcal/mol$, and that the 
next higher local minimum has an energy of $E_1 = -9.8~kcal/mol$.\cite{EH96d}
Hence, we identify any configuration with energy below $E=-9.8~kcal/mol$
as a representative of the ground state.

In the ELP simulations of Met-enkephalin, we used the potential energy itself 
as an order parameter and thus the deformed energy landscape is generated by
$\tilde{E} = E + H(E,t)$,
where $H(E,t)$ is the  histogram in energy at MC sweep $t$. 
We chose a bin size $E_{bin} = 0.25~kcal/mol$ in the histogram, but checked
our results also for $E_{bin} = 0.5~kcal/mol$ and $E_{bin} = 1~kcal/mol$
without finding noticeable differences in our results. Setting the temperature
to $T=50~K$, and with $\beta = 1/k_BT$ we find as a weight for the
MC simulation:
\begin{equation}
w(E,t) = e^{-\beta(E+H(E,T))}~.
\end{equation}

The characteristic behavior of our ELP method is  exemplified in
Fig.~1 which shows for Met-enkephalin the time series of a simulation 
with 50,000 sweeps. The starting configuration has an energy of
$E_{start}=-5.1~kcal/mol$ and was obtained from a random configuration
through quenching in initial 100 sweeps. The simulation soon gets
trapped in a local minimum of $E\approx -7.8~kcal/mol$ (after only 250
MC sweeps). Through the following MC sweeps entries in the corresponding
histogram bin are accumulated and  the energy landscape locally deformed, 
until  after about 750 MC sweeps the simulation escapes
this local minimum to find a lower local minimum after 2000 MC sweeps.
This process is repeated till the simulation finds the global minimum
conformation for the first time after 7260 sweeps. Within the 50,000
sweeps of our simulation the ground state region ($E < -9.8~kcal/mol$)
was visited 5 times, each visit separated by explorations in the
high energy region. 

Note that the range of energies covered increases
with MC time: ELP starts with filling up the small
`potholes' in the energy landscape, but later in the
simulation  large valleys are also filled up. Hence, our algorithm 
is self-adjusting: with increasing length of the simulation it becomes
possible to overcome higher and higher energy barriers. In that regard,
ELP is more efficient than standard techniques such as
simulated annealing \cite{SA} where the height of energy barriers that can be 
overcome shrinks with MC time. In order to test that conjecture
we performed 20 independent 
runs  of 50,000 MC sweeps with ELP and
compared the results with  20 simulated annealing runs of equal
statistics. 
In the simulated annealing runs the temperature was lowered
exponentially in 50,000 sweeps from an initial temperature $T=1000~K$
to a final temperature $T=50~K$. In Ref.~\onlinecite{HO94c} this proved to be
the optimal annealing schedule for Met-enkephalin.  Even with
this optimized annealing schedule,  the ground state was  found only in 
$8/20 = 40\%$ of the runs (in an average time of 43,000 MC sweeps)
and the average value of the 
lowest energy conformation ($<E_{min}>=-8.5~ kcal/mol$) was above our 
threshold for ground state configurations ($-9.8~kcal/mol$) and subject
to large fluctuations (with standard deviation $\sigma=2.1~kcal/mol$).
Better results were obtained in 20 tabu search runs with same 
statistics. Here the ground state region was found in $10/20 = 50\%$ of
the runs  and the average value of the lowest energy conformations was
$<E_{min}> = -9.5 kcal/mol$. 
On the other hand, with ELP we found the ground state in each 
of the 20 runs (in, on average, 12,700 MC sweeps) and
the average of lowest energy states 
$<E_{min}> = -10.3~kcal/mol$ was well
below our threshold for ground state configuration, subject to only
small fluctuations ($\sigma=0.3~kcal/mol$). 

In order to further evaluate  the ELP approach 
we also studied  the much larger villin headpiece subdomain, 
36-residue peptide (HP-36) that  
is with 597 atoms about 8 times larger than Met-enkephalin (75 atoms).
HP-36 is one of the smallest peptides that can fold
autonomously and was chosen recently for
a 1-microsecond molecular dynamics simulation of protein folding.\cite{Kol} 
The experimental structure was determined by NMR analyses.\cite{McKnight1}   
Since it is a
solvated molecule we also had to take into account the interaction
between protein and solvent.  We have approximated this
contribution to the overall energy by adding a solvent accessible surface 
term \cite{sur_solvent} to the energy function:
$E = E_{Ecepp/2} + \sum_i \sigma_i A_i~.$
Here, the sum goes over all atoms and the $A_i$ are the 
solvent accessible surface areas of the atoms. The
parameters $\sigma_i$ were chosen from Ref.~\cite{SET4}.

HP-36 allows in a simple way the definition of an order parameter to
characterize configurations other than by their energy. This natural order
parameter is the number $n_H$ of residues in the peptide which are part of an
$\alpha-$helix. Following earlier work \cite{OH95} we define a residue
as helical if the pair of backbone dihedral angles $\phi,\psi$ takes a
value in the range $(-70\pm 20,-37\pm 20)$. Throughout the search process 
we tried now  to deform the energy landscape 
by means of a histogram $H(E,n_H,t)$ in {\it both} helicity and energy: 
$\tilde{E}  =  E + H(E,n_H,t)$.  Operating again
at a temperature $T=50$ K, we find as weights for the search algorithm
\begin{equation}
w(E,n_H,t) = e^{-\beta (E+H(E,n_H,t))}~.
\end{equation}
Using this weight we performed simulations with 50,000 MC sweeps
(starting from random configurations)  keeping track of the lowest 
energy configuration during the search process. 

The structure of HP-36 as obtained from the Protein Data
Bank (PDB  code 1vii) is shown in Fig.~2. The structure consists 
of three helices between
residues 4-8, 15-18, and 23-32, respectively, which are connected by a 
loop and a turn. After regularizing this structure with the 
program FANTOM \cite{FANTOM} we obtained as its   energy (ECEPP/2 + solvation
term)  $E_{nat} = -276$ kcal/mol.  Our new ELP method led after 25,712
MC sweeps to a  configuration with lowest energy $E_{min} = -277$ 
kcal/mol which we show in Fig.~3. The above structure has a radius 
of gyration $R_{\gamma} = 10.1$ \AA  ~indicating that the 
numerically obtained structure is slightly less 
compact than the  experimental structure ($R_{\gamma}=9.6$\AA). 
 It consists of three helices where the first helix 
stretches from residue 2 to residue 11 and is more elongated than the
corresponding one in the native structure (residues 4-8). The second helix
consist of residues 13-17 (compared to residue 15-18 in the native structure)
and the third helix stretches from residue 23-33 (residues 23-32 in the PDB
structure). The structure has 95\% of the native helical content, that is
95\% of all residues which are part of a helix in the experimental structure
are also part of a helix in our structure. We also note that 65\% of the
native contacts were formed in our structure (two residues $i$ and $j$ 
($j > i+2$) are taken to be in contact if their $C_\alpha$ atoms are closer 
than 8.5 \AA). Both values are comparable with the results in 
Ref.~\onlinecite{Kol} (but required orders of magnitude less
computer time) where the optimal structure of a $1~\mu s$ molecular 
dynamic folding simulation showed 80\% of native helical content and
62 \% of native contacts. Similarly comparable were the values of the
root-mean-square deviation (RMSD) of both numerically determined
conformers to the native structure: $5.8 $ \AA~versus $5.7 $ \AA~in 
Ref.~\onlinecite{Kol} when all backbone atoms where counted. 

We conclude that  even for large peptides such as HP-36 
our novel optimization method is able to find structures that 
are close to the experimentally determined ones. In passing, we remark
that an exploratory simulated annealing  run of 100,000 sweeps
did not lead to such structures.  However, our ELP prediction
of the HP-36 structure is  limited to an RMSD of $\approx 6$ \AA.
This points to a general problem in protein simulations:
it is not clear whether the utilized cost function has indeed
the biologically active structure of a given protein as its global minimum.
In fact, our optimal structure has slightly lower energy than the
native one. The problem becomes obvious when solvation effects are
neglected. An ELP run of 50,000 sweeps  relying only on the 
ECEPP/2 force field led to a lowest-energy structure 
with an ECEPP energy of $E_{GP} = - 192$ kcal/mol (found after 25,712
sweeps). That structure,
build out of two helices (between residues 2-16 and 23-33) 
connected by a loop,  differs significantly from the  regularized 
PDB-structure with the higher potential energy $E_{nat} = -176$ kcal/mol. 
Hence, the native structure of the peptide HP-36 is {\it not} the 
global minimum configuration in ECEPP/2. Only  the inclusion of the
solvation term led to an essentially correct structure as global minimum
configuration.

Summarizing, we have developed a new and general 
stochastic global optimization method  that is easy to implement and 
combines energy landscape deformation ideas with elements of tabu search.  
The efficiency of ELP was compared with  simulated 
annealing \cite{SA} and  tabu search. \cite{glover,cvijovic}
To illustrate the power of our novel approach, we
applied it to the structure prediction of  HP-36, a 36 residue
peptide. For this large peptide an unbiased all-atom simulation 
using ELP led to a 3D structure 
very close to the experimentally determined one. In future
work we want to  extend application of our new approach to other
optimization problems. \cite{HLH01e}

\noindent
{\bf Acknowledgement}\\
U.H.E. Hansmann  acknowledges support by a  research grant
of the National Science Foundation (CHE-9981874) and 
by NSF MRI grant \#9871133. Part of this article 
was written while U.H. was visitor at the Department of Physics at
the Bielefeld University. He thanks F.~Karsch for his hospitality
during his stay in Bielefeld.

\newpage
\noindent
{\bf \Large Figures:}
\begin{enumerate}
\item Time series of a minimization run of 50,000 sweeps for
      the pentapeptide Met-enkephalin.
\item NMR derived structure of the 36 residue peptide HP-36 as obtained from
      the PDB data base (1vii).
\item Lowest-energy structure of HP-36 as obtained with ELP.
\end{enumerate}

\newpage
\cleardoublepage

\begin{figure}[t]
\begin{center}
\begin{minipage}[t]{0.95\textwidth}
\centering
\includegraphics[angle=-90,width=0.72\textwidth]{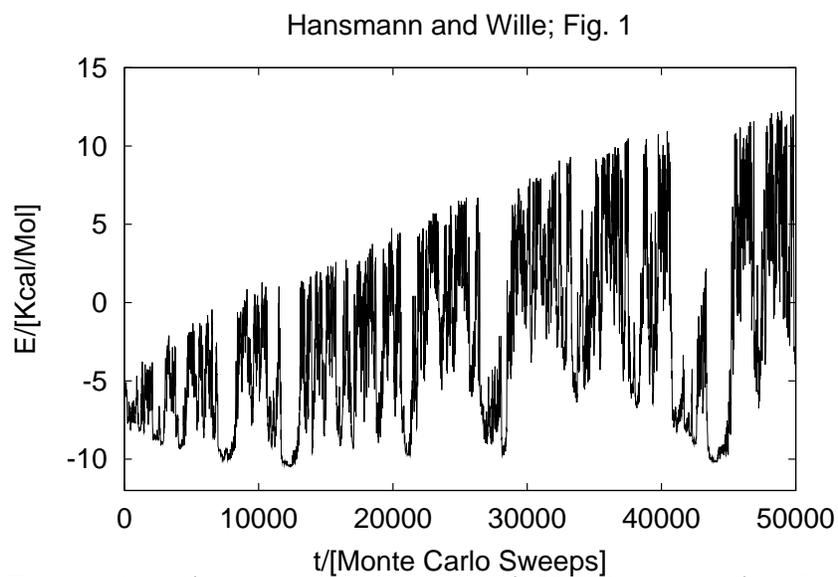}
\renewcommand{\figurename}{FIG.}
\caption{Time series of a minimization run of 50,000 sweeps 
for the pentapeptide Met-enkephalin.}
\label{fig1}
\end{minipage}
\end{center}
\end{figure}

\newpage
\cleardoublepage
\begin{figure}[!ht]
\begin{center}
\begin{minipage}[t]{0.95\textwidth}
\centering
\includegraphics[angle=-90,width=0.72\textwidth]{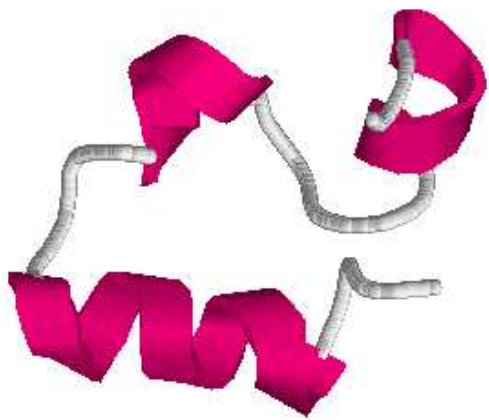}
\renewcommand{\figurename}{FIG.}
\caption{NMR derived structure of the 36 residue peptide HP-36 
          as obtained from the PDB data base (1vii).}
\label{fig2}
\end{minipage}
\end{center}
\end{figure}

\newpage
\cleardoublepage
\begin{figure}[!ht]
\begin{center}
\begin{minipage}[t]{0.95\textwidth}
\centering
\includegraphics[angle=-90,width=0.72\textwidth]{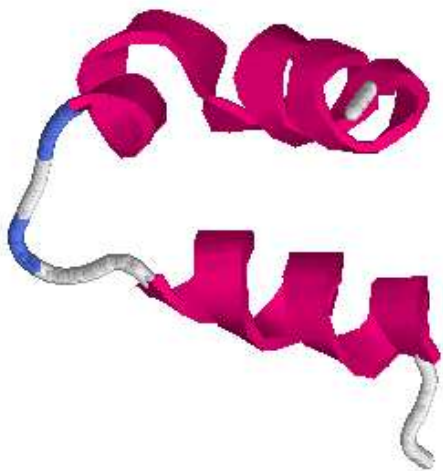}
\renewcommand{\figurename}{FIG.}
\caption{Lowest-energy structure of HP-36 as obtained with ELP.}
\label{fig3}
\end{minipage}
\end{center}
\end{figure}
\end{document}